\documentclass[10pt]{article}
\usepackage{amsfonts}
\usepackage{amsmath}
\usepackage{amssymb}
\usepackage{epsfig}
\usepackage{graphicx,color}
\usepackage{graphicx}
\usepackage{mathptmx}
\newcommand{\spur}[1]{\not\! #1 \,}
\begin{document}
\title{Annihilation-Type Charmless Radiative\\ Decays of $B$ Meson in  Non-universal $Z^\prime$ Model }
\author{~~Juan Hua$^{1}$\footnote{Email:juanhua@126.com},
~~C.~~S. Kim$^{2}$\footnote{Email:cskim@yonsei.ac.kr},~~ Ying
Li$^{1,2}$\footnote{Email:liying@ytu.edu.cn} \\
 \small {\it  1. Department
of Physics, Yantai University, Yantai 264-005, China } \\
\small{\it 2.Department of Physics and IPAP, Yonsei University, Seoul
120-479, Korea }\\
} \maketitle
\begin{abstract}
\noindent We study charmless pure annihilation type radiative
$B$ decays within the QCD factorization approach. After adding the
vertex corrections to the naive factorization approach, we find that  the branching ratios of
$\overline{B}^0_d\to\phi\gamma$, $\overline{B}^0_s\to\rho^0\gamma$
and $\overline{B}^0_s\to\omega\gamma$ within the standard model are at the order of
$\mathcal{O}(10^{-12})$, $\mathcal{O}(10^{-10})$ and
$\mathcal{O}(10^{-11})$, respectively. The smallness of these decays
in the standard model makes them sensitive probes of flavor physics
beyond the standard model. To explore their physics potential, we have
estimated the contribution of $Z^\prime$ boson in the decays. Within
the allowed parameter space, the branching ratios of these decay modes can
be enhanced remarkably in the non-universal $Z^\prime$ model:
The branching ratios can reach to $\mathcal{O}(10^{-8})$ for
$\overline B_s^0 \to \rho^0(\omega)\gamma$ and
$\mathcal{O}(10^{-10})$ for the $\overline B_d^0 \to \phi \gamma$,
which  are large enough for LHC-b and/or Super B-factories to detect
those channels in near future. Moreover, we also predict large CP
asymmetries in suitable parameter space. The observation of these
modes could in turn help us to constrain the $Z'$ mass within the model.
\end{abstract}
\newpage
\section{Introduction}
Rare $B$ decays induced by flavor changing neutral currents (FCNC)
play important roles in particle physics, where they are
always regarded as ideal places for probing signals of  new
physics. The GIM suppression of FCNC amplitude is absent in many new
physics scenarios beyond the standard model (SM), which could give
large enhancement of FCNC contributions over the SM predictions.
However, due to our poor knowledge of non-perturbative QCD,
predictions for many interesting exclusive decays are polluted by
large hadronic uncertainties. Therefore, it would be of great interest to
explore rare $B$ decays, which are induced with few hadronic parameters as well
as only by FCNC currents. Two body radiative $B$ decays
involve simple hadronic dynamics with only one hadron in the final
states, so they suffer much less pollution than non-leptonic decays.

In studying the radiative decays such as $B\to K^* \gamma$,
$\rho (\omega) \gamma$, the isospin breaking
effects between the charged $B^\pm$ and neutral $B^0$  in these
modes are mainly from the annihilation type diagrams
\cite{Ali:2006fa,Bosch:2001gv,Ball:2006eu,Wang:2007an}. Many of
pure annihilation type radiative decays, such as $B\to \phi \gamma$
and $B\to J/\psi \gamma$, have been analyzed in the QCD factorization
approach \cite{Li:2003kz,Lu:2003ix} and in the perturbative QCD approach
\cite{Li:2006xe}. We find the branching ratio of $B\to\phi\gamma$ is
at the order of $\mathcal{O}(10^{-11}\sim10^{-12})$ in the SM. The
decay rate is too small to be observed at presently running $B$ factories,
BaBar and Belle. Any measurements of the decays  at BaBar and Belle
would be direct evidences of new physics. In this work, we
explore the decay $B\to \phi \gamma$ and similarly the decay
$\overline B_s^0\to \rho^0 (\omega) \gamma$ in the non-universal
$Z^\prime$ model \cite{Langacker:2000ju}, which could be naturally
derived in certain string constructions \cite{Buchalla:1995dp}, E6
models \cite{Nardi:1992nq} and so on. Generally speaking, within the such  model a flavor mixing can
be induced at the tree level in the up-type and/or down-type quark
sector after diagonalizing their mass matrices. In some new physics
model, FCNCs due to $Z^\prime$ exchange can be induced by mixing
among the SM quarks and the exotic quarks, which have been predicted to have
different $Z^\prime$ quantum numbers. Here we will consider the
model in which the interaction between the $Z^\prime$ boson and
fermions are flavor non-universal for left handed couplings and
flavor diagonal for right handed couplings. The effects of the
$Z^\prime$ on  other processes of the interest have been
investigated in a number of papers such as
\cite{Erler:2009jh,Chen:2008za}, especially in $B$ physics
\cite{Barger:2009hn,Cheung:2006tm,Chang:2009wt,Chen:2009bj}. The
recent review about $Z^\prime$ in detail is referred to Ref.
\cite{Langacker:2008yv}.

To keep completeness, we first calculate these decays in the naive
factorization approach. Then we  add the vertex
corrections to the four quark operators, which have been performed
in the so called QCD factorization approach \cite{Beneke:1999br} in the SM,
utilizing the light-cone wave functions of the light vector mesons.
A similar work within the R-parity violating SUSY can be also found in Ref. \cite{Li:2003kz}. However, in
this work we will revisit these processes with the updated parameters in the non-universal $Z^\prime$ model.

\section{Calculation in the Standard Model}
In the SM, the common starting point is the effective weak
Hamiltonian which mediates flavor-changing neutral current
transitions of the type $b\to D~(D=d,s) $ :
\begin{eqnarray}\label{hamiton}
 {\cal H}_{eff}={G_F\over \sqrt 2}\Big[\sum\limits_{p=u,c}V_{pb}V^*_{pD}\Big(
 C_1O_1^p+C_2O_2^p\Big)-V_{tb}V^*_{tD}\sum\limits_{i=3}^{10,7\gamma,8g}
 C_iO_i\Big].
\end{eqnarray}
The explicit forms of the operators $O_i$  and the corresponding
Wilson coefficients $C_i$ at the scale of  $\mu=m_b$ can be found in
Ref. \cite{Buras}. $V_{p(t)b}$, $V_{p(t)D}$ are the
Cabibbo-Kabayashi-Maskawa (CKM) matrix elements. According to the
effective Hamiltonian (\ref{hamiton}), we can draw the lowest order
diagram of this channel, as shown in Fig. 1, which is dominated by the
photon radiated from light quark in  $\overline B_s^0$
meson. When the photon is radiated from heavy $b$ quark, the
energetic up or down quark will suppress the $\rho^0$  production by a
power of $\Lambda_{QCD}^2/m^2_b$, therefore, here we neglect its contribution.
This point have been  clearly  discussed in $B\to K^* \gamma$
decays \cite{Bosch:2001gv,Ball:2006eu}.

\begin{figure}
\begin{center}\label{fig:1}
\scalebox{0.7}{\epsfig{file=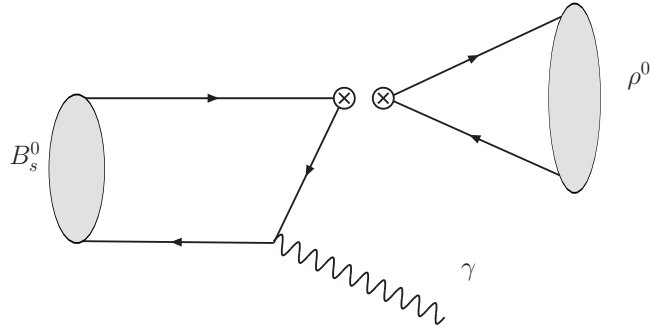}} \end{center} \caption{\small
The leading diagram  for $\overline{B}^0_s \to\rho^0\gamma$. The
symbol $\otimes$ denotes the insertion of penguin operators $O_i$. }
\end{figure}

For convenience, we denote that $\eta^{*}_{\perp}$ and
$\epsilon^{*}_{\perp}$ are transverse polarization vectors of the
final vector meson and photon, respectively. The photon energy and
momentum are defined as $E_\gamma$ and $q$, and the momentum of
$\overline B_q$ meson is $p_B=mv$. In the rest frame of $\overline B_q$ meson,
we take the photon and the vector meson moving along the
$n_{-}=(1,0,0,-1)$ and $n_{+}=(1,0,0,1)$ directions, respectively.
Within the effective Hamiltonian and naive factorization hypothesis,
we can write down the amplitudes as follows:
\begin{multline}
A(\overline {B^{0}_s} \to\rho^0\gamma)=\frac{G_{F}}{2}\left[V_{ub}
V_{us}^*a2-V_{tb} V_{ts}^* \left(\frac{3}{2} a_7+\frac{3}{2} a_9
\right)
    \right] \\
\times \sqrt{4\pi\alpha_e}f_{\rho}m_{\rho} F_V \biggl\{
-\epsilon_{\mu\nu\rho\sigma}
\eta^{*\mu}_{\perp}\epsilon^{*\nu}_{\perp} v^{\rho}q^{\sigma}
+i\left[(\eta_{\perp}^{*}\cdot\epsilon^{*}_{\perp}) ( v\cdot
q)-(\eta^{*}_{\perp}\cdot q)
(v\cdot\epsilon^{*}_{\perp})\right]\biggr\},
\end{multline}
\begin{multline}
A(\overline {B^{0}_s} \to\omega\gamma)=\frac{G_{F}}{2}\left[V_{ub}
V_{us}^*a2-V_{tb} V_{ts}^* \left(2a_3+2 a_5
+\frac{1}{2}C_7+\frac{1}{2} a_9 \right)
    \right] \\
\times \sqrt{4\pi\alpha_e}f_{\rho}m_{\rho} F_V \biggl\{
-\epsilon_{\mu\nu\rho\sigma}
\eta^{*\mu}_{\perp}\epsilon^{*\nu}_{\perp} v^{\rho}q^{\sigma}
+i\left[(\eta_{\perp}^{*}\cdot\epsilon^{*}_{\perp}) ( v\cdot
q)-(\eta^{*}_{\perp}\cdot q)
(v\cdot\epsilon^{*}_{\perp})\right]\biggr\},
\end{multline}
\begin{multline}
A(\overline {B^{0}_d}
\to\phi\gamma)=\frac{G_{F}}{\sqrt{2}}\left[-V_{tb} V_{td}^*
\left(a_3+a_5-\frac{1}{2} a_7 -\frac{1}{2}a_9 \right)
    \right] \\
\times \sqrt{4\pi\alpha_e}f_{\rho}m_{\rho} F_V \biggl\{
-\epsilon_{\mu\nu\rho\sigma}
\eta^{*\mu}_{\perp}\epsilon^{*\nu}_{\perp} v^{\rho}q^{\sigma}
+i\left[(\eta_{\perp}^{*}\cdot\epsilon^{*}_{\perp}) ( v\cdot
q)-(\eta^{*}_{\perp}\cdot q)
(v\cdot\epsilon^{*}_{\perp})\right]\biggr\},
\end{multline}
where $a_i$ is defined as the combination of the Wilson
coefficients,
\begin{eqnarray}\label{a1}
 a_i=C_i+\frac{C_{i\pm1}}{N_c},
\end{eqnarray}
for an odd (even) value of $i$. The form factor $F_V$ has been
defined in  Refs.
\cite{Lunghi:2002ju,DescotesGenon:2002mw,Ball:2003fq}
\begin{multline}
\langle \gamma(\epsilon^{*},q){\mid} \bar{d}\gamma_{\mu}
(1-\gamma_{5})b {\mid}\bar{B}^0_d(v) \rangle =
\sqrt{4\pi\alpha_e}\Big [ -F_{V}(E_{\gamma})
\epsilon_{\mu\nu\rho\sigma}\epsilon^{*\nu}v^{\rho}q^{\sigma}
+iF_{A}(E_{\gamma})(\epsilon^{*}_{\mu}
q{\cdot}v-q_{\mu}v{\cdot}\epsilon^{*}) \Big ].
\end{multline}
In order to calculate the form factor $F_V$, we need
two-particle light-cone projector for an initial $B$ meson:
\begin{eqnarray}
{\cal M}^B_{\alpha\beta}&=& \frac{i}{4 N_c}f_{B}M_{B}
\biggl\{(1+\spur{v})\gamma_5 \Big
[\Phi_{B_1}(l_{+})+\spur{n_{-}}\Phi_{B_2}(l_{+}) \Big]
\biggr\}_{\alpha\beta},
\end{eqnarray}
where $\Phi_{B_1}(l_{+})$ and $\Phi_{B_2}(l_{+})$ are the leading
twist light-cone distribution functions \cite{Grozin:1996pq}. Thus,
we obtain the standard result:
\begin{equation}
F_{V}(E_{\gamma})=F_{A}(E_{\gamma})=\frac{Q_s f_{B}M_B}{
2\sqrt{2}E_{\gamma} }\int dl_{+}\frac{\Phi_{B_1}(l_{+})}{l_{+}},
\end{equation}
where $Q_s=-1/3$ is the charge of the $s$ quark in units of the
proton's charge. Because we have little knowledge about the
distribution of the heavy meson,  the integral in above formulae
is often parameterized as :
\begin{equation}
\int dl_{+}\frac{\Phi_{B_1}(l_{+})}{l_{+}}=\frac{1}{\lambda_B}.
\end{equation}
Consequently, we write down the helicity amplitudes for these
channels as:
\begin{eqnarray}
{\cal M}_{B_s\to \rho \gamma}^{+
+}&=&i\frac{G_F}{2}\sqrt{4\pi\alpha_e}F_{V}
f_{\phi}m_{\phi}M_{B}\left[V_{ub} V_{us}^*a_2-V_{tb} V_{ts}^*
\left(\frac{3}{2} a_7 +\frac{3}{2} a_9 \right)\right],\\
{\cal M}_{B_s\to \omega \gamma}^{+
+}&=&i\frac{G_F}{2}\sqrt{4\pi\alpha_e}F_{V}
f_{\phi}m_{\phi}M_{B}\left[V_{ub} V_{us}^*a2-V_{tb} V_{ts}^*
\left(2a_3+2 a_5
+\frac{1}{2}C_7+\frac{1}{2} a_9 \right)\right],\\
{\cal M}_{B_d\to \phi \gamma}^{+
+}&=&i\frac{G_F}{\sqrt{2}}\sqrt{4\pi\alpha_e}F_{V}
f_{\phi}m_{\phi}M_{B}\left[-V_{tb} V_{td}^*
\left(a_3+a_5-\frac{1}{2} a_7 -\frac{1}{2}a_9 \right)\right],\\
{\cal M}_{B_s\to \rho \gamma}^{--}&=&{\cal M}_{B_s\to \omega
\gamma}^{--}={\cal M}_{B_d\to \phi \gamma}^{--}=0.
\end{eqnarray}

\begin{table}
\begin{center}
\caption{Summary of input parameters} \label{parameter}
\begin{tabular}{c}\hline\hline
 $\begin{array}{cccccccccc}
 \lambda & A &\bar{\rho}& \bar{\eta}&\Lambda_{\overline{\mathrm{MS}}}^{(f=4)}
 &\tau_{B^0}&\tau_{B_s^0}&\lambda_B&\alpha_e&\alpha_s\\
 0.225& 0.818&0.141& 0.348& 250
 \mbox{MeV}&1.54\mbox{ps}&1.46\mbox{ps}&0.35&1/132 &0.214
 \end{array}$\\
 \hline
 $\begin{array}{cccccccc}
   f_B &f_{B_s} & f_{\rho} &f_{\rho}^\perp & f_{\phi} &f_{\phi}^\perp & f_{\omega} &f_{\omega}^\perp\\
  216\mbox{MeV}&236\mbox{MeV}& 210\mbox{MeV} & 150\mbox{MeV}&221\mbox{MeV}
  &175\mbox{MeV}& 187\mbox{MeV} & 151\mbox{MeV}
 \end{array}$ \\\hline
 $\begin{array}{ccccc}
     m_B&  m_{B_s}  & m_{\phi} & m_{\rho}& m_{\omega}\\
 5.28\mbox{GeV} &  5.36\mbox{GeV}&1.02\mbox{GeV}&0.77\mbox{GeV}&0.78\mbox{GeV}
 \end{array}$ \\
   \hline\hline
\end{tabular}
\end{center}
\end{table}

Depending on the parameter values  listed in Table 1, one can
get the averaged branching ratios as:
\begin{eqnarray}
{\cal B}(\overline {B}^0_s \to\rho^0\gamma )=1.1\times 10^{-10}; \nonumber\\
{\cal B}(\overline {B}^0_s \to\omega\gamma )=5.6\times 10^{-11};\nonumber\\
{\cal B}(\overline {B}^0_d \to\phi^0\gamma )=1.7\times10^{-13}.
\label{r1}
\end{eqnarray}
Within the naive factorization hypothesis, because of no strong phases entering
into these processes,  there should not exist  any CP asymmetry for the processes.

\begin{figure}
\begin{center}
\scalebox{0.7}{\epsfig{file=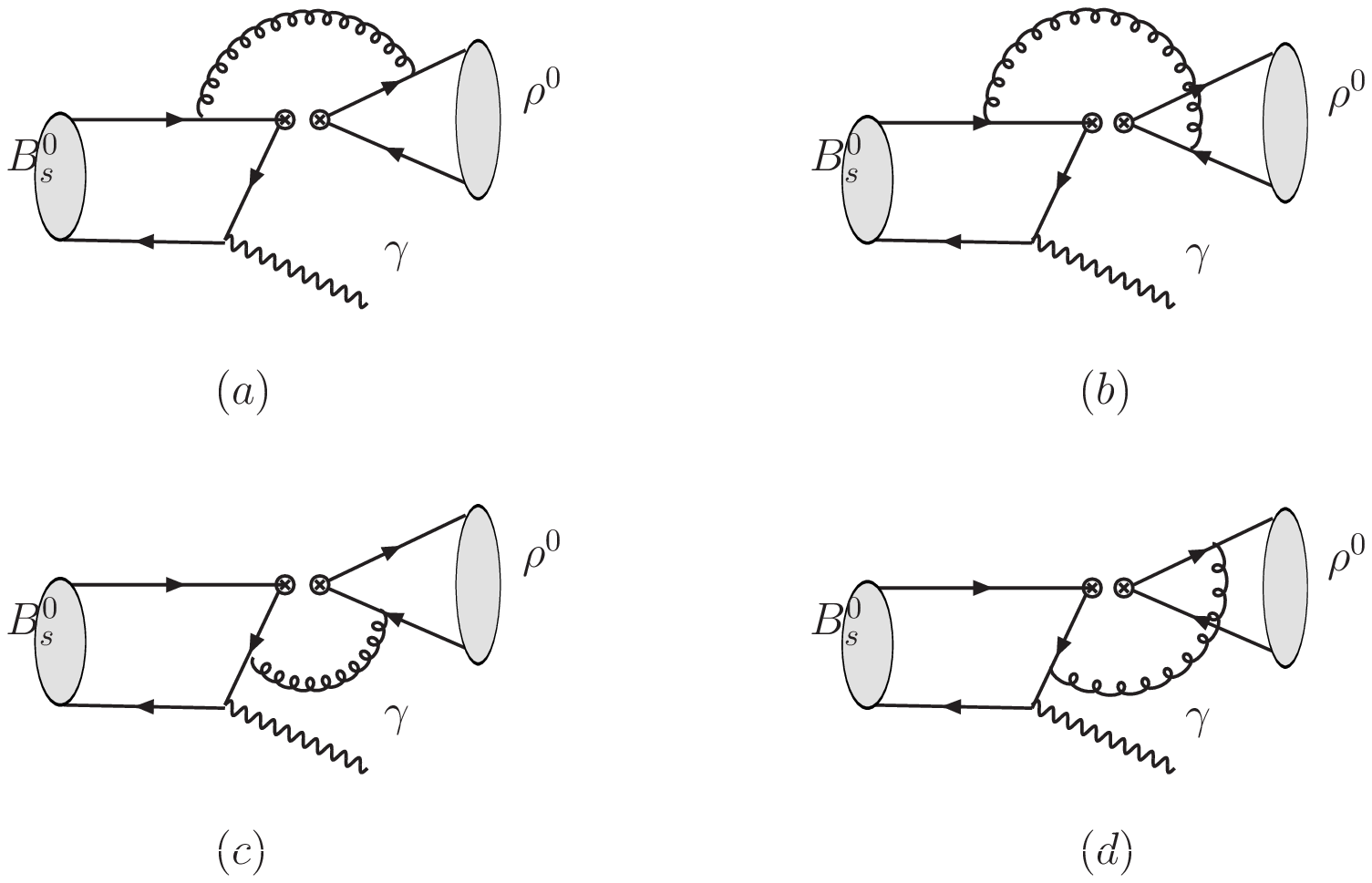}}
\end{center}\label{fig2}
\caption{\small Non-factorizable diagrams for $\overline{B}^0_s
\to\rho^0\gamma$. }
\end{figure}

Up to now in our calculation, non-factorizable contributions have been
neglected. As next step, we add the vertex corrections and
the leading non-factorizable diagrams, shown in Fig. 2. To achieve
the goal, the QCD factorization framework
\cite{Beneke:1999br} proposed by Beneke, Buchalla, Neubert and
Sachrajda is very suitable to be applied. To calculate non-factorizable
diagrams, we also need the two-particle
light-cone projector of the vector mesons:
\begin{eqnarray}
 {\cal M}^{\rho}_{\perp\alpha\beta}=-\frac{f^{\perp}_{\rho}m_{\rho}}{4N_c}
\biggl\{\spur{\epsilon}^*_{\perp}g^{(v)}_{\perp}(u)+\frac{i}{8}\epsilon_{\mu\nu\rho\sigma}
\epsilon^{*\nu}_{\perp}n^{\rho}_{+}n^{\sigma}_{-}
\gamma^{\mu}\gamma_{5}\frac{\partial g^{(a)}_{\perp}(u)}{\partial
u}\biggr\}_{\alpha\beta},
\end{eqnarray}
where $g^{(v)}_{\perp}(u)$ and $g^{(a)}_{\perp}(u)$ are twist-3
distribution amplitudes of vector mesons, and explicit formulae can
be found in Ref. \cite{Ball:1998sk}.

After adding the contributions, the form of amplitudes for the decay
modes becomes the similar, just replacing $a_i$ by
$a_i^\prime$, which involve the ${\cal O}(\alpha_s )$ corrections.
$a_{i}^{\prime}$'s are calculated to be
\begin{eqnarray}\label{a2}
a^{\prime}_2&=&a_2
+\frac{\alpha_s}{2\pi}\frac{C_F}{N_C}\frac{f_{\rho}^{\perp}}{f_{\rho}}C_{1}F_{1},\nonumber\\
a^{\prime}_3&=&a_3
+\frac{\alpha_s}{4\pi}\frac{C_F}{N_C}\frac{f_{\rho}^{\perp}}{f_{\rho}}C_{4}F_{1},\nonumber\\
a^{\prime}_5&=&a_5
+\frac{\alpha_s}{4\pi}\frac{C_F}{N_C}\frac{f_{\rho}^{\perp}}{f_{\rho}}C_{6}F_{2},\nonumber\\
a^{\prime}_7&=&a_7
+\frac{\alpha_s}{4\pi}\frac{C_F}{N_C}\frac{f_{\rho}^{\perp}}{f_{\rho}}C_{8}F_{2},\nonumber\\
a^{\prime}_9&=&a_9
+\frac{\alpha_s}{4\pi}\frac{C_F}{N_C}\frac{f_{\rho}^{\perp}}{f_{\rho}}C_{10}F_{1},
\end{eqnarray}
where $F_{1,2}$ arise  from one gluon exchange between the two
currents of color-octet operators as shown in Fig. 2,
\begin{eqnarray}
F_{1}&=&\int^1_0 du\left(\frac{g_{\perp}^{(a)\prime}(u)}{4}
-g^{(v)}_{\perp}(u) \right) \left[ -14-3i\pi-12\ln\frac{\mu}{m_b}
\right. \nonumber\\ &&\left. +\left( 5+\frac{u}{1-u}\right)\ln u
-\frac{\pi^2}{3}+2\mathrm{Li}_{2}(\frac{u-1}{u})\right],\\
F_{2}&=&\int^1_0 du\left(
g^{(v)}_{\perp}(u)+\frac{g_{\perp}^{(a)\prime}(u)}{4}
                   \right)
\left[ -14-3i\pi-12\ln\frac{\mu}{m_b} \right. \nonumber\\ &&\left.
+\left( 5+\frac{u}{1-u}\right)\ln u
-\frac{\pi^2}{3}+2\mathrm{Li}_{2}(\frac{u-1}{u})\right].
\end{eqnarray}
Here we  have neglected the small effect of box diagrams and the
diagrams with photon  radiating from energetic light quarks, which
are further suppressed by $\Lambda_{QCD}/M_{B}$. Including ${\cal
O}(\alpha_{s})$ contributions, the averaged branching ratios in the SM are
estimated to be
\begin{eqnarray}
{\cal B}(B_s^0\to\rho^0\gamma )&=&1.1\times 10^{-10},\nonumber \\
{\cal B}(B_s^0 \to\omega\gamma )&=&2.3\times 10^{-11},\nonumber \\
{\cal B}(B_d^0 \to\phi\gamma )&=&2.9\times 10^{-12}.
\end{eqnarray}
Comparing with the results in Eq. (\ref{r1}) of the naive
factorization, one finds the branching ratio of
$B_s^0\to\rho^0\gamma$ almost unchanged. To find out the reason why the
correction dose not take an effect, we list the values of
$a_i^{\prime}$ of these decay modes in the Table 2. From the table, we
find that the corrections to $a_{7,9}$ are very small and can be
neglected. Although Wilson coefficients of QCD penguin operators
changed a little, but they give no contribution because the quark
component of $\rho^0$ is $(u\bar u- d\bar d)/\sqrt{2}$. For $a_2$,
it changes much, but the correction is suppressed by the CKM
elements. So, the unchanged branching ratio is quite reasonable. As for
$B_s^0\to\omega\gamma$, the decrease of $a_5$ can cause that the
branching ratio becomes even smaller than that of the naive
factorization. For the decay $B_d^0\to\phi\gamma$, the increase of
the ratio mainly comes from the change of $a_3$ and $a_5$.

\begin{table}
\begin{center}\caption{The values of $a_i$ in different scenario. In the LO column, $a_i$ are defined in
Eq. (\ref{a1}); in the NLO column, the values are $a_i^\prime$
defined in Eq. (\ref{a2}); in the $Z^{\prime}$ Model, they are values
of  $a_i^\prime+ \Delta a_i^\prime$, which are defined
in Eq. (\ref{a3}).}
\begin{tabular}{|c|c|c|c|}
\hline\
$a_i^{(\prime)}+(\Delta a_i^{(\prime)})$ &  LO &  NLO  & $Z^{\prime}$ Model \\
  \hline
  $a_2$ &  $0.170$ & $0.149 - i 0.010$  & $0.149-i 0.010  $ \\
  $a_3$ &  $0.002$ & $-0.002 - i 0.002$ &$(0.019+i 0.007 )   \xi e^{i \phi }+(-0.002 - i 0.002) $ \\
  $a_5$ & $-0.005$& $0.003 + i 0.002$  &$(0.009-i 0.008 )   \xi e^{i \phi }+(0.003 + i 0.002)  $ \\
  $a_7$ & $ 0.000$ & $-0.000 + i 0.000$  & $(3.798-i 0.067 ) \xi e^{i \phi }+(-0.000 + i 0.000) $ \\
  $a_9$ & $-0.008$ & $-0.008 - i 0.000$  & $(3.932-i 0.050 ) \xi e^{i \phi }+(-0.008 - i 0.000) $ \\
  \hline
\end{tabular}
\end{center}
\end{table}

Because there are both weak  and  strong phases in the decay
modes $B_s^0 \to\rho^0\gamma$ and $B_s^0 \to\omega\gamma$, we can
get the CP asymmetries of these two channels as follows,
\begin{eqnarray}
&&{\cal A}(B_s^0 \to\rho\gamma )=3\%;\nonumber \\
&&{\cal A}(B_s^0 \to\omega\gamma )=-27\%,
\end{eqnarray}
by the definition of CP asymmetry
\begin{eqnarray}
{\cal A}=\frac{ BR(\overline {B_s^0} \to V\gamma)- BR(B_s^0 \to
V\gamma)}{BR(\overline {B_s^0} \to V\gamma)+ BR(B_s^0 \to V\gamma)}.
\end{eqnarray}
For the decay mode  $B_d^0\to \phi\gamma$, there is only weak
phase from  $V_{tb}V_{td}^*$, so that the CP asymmetry in this
decay disappears within the SM.

\section{Calculation in the Non-universal $Z^\prime$ Model}

Now we  consider the effects due to an extra $U(1)^\prime$ gauge
boson $Z^\prime$. Usually, the flavor mixing can be induced at the tree
level in up-type and/or down-type quark sector after
diagonalizing their mass matrices. In some new physics model, FCNCs
due to $Z^\prime$ exchange can be induced by mixing among the SM
quarks and the exotic quarks, which is predicted and can make different
$Z^\prime$ quantum numbers after the mixing. Here we will consider the model in which
the interaction between the $Z^\prime$ boson and fermions are flavor
non-universal for left handed couplings and flavor diagonal for
right handed couplings. For simplicity, we neglected the mixing
between the $Z^0$ and $Z^\prime$ and the evolution effect from the
high scale $M_{Z^\prime}$ to the $M_W$ scale.

We start to set up the relevant interactions with the new $Z^\prime$
gauge particle. Following the convention in Ref.
\cite{Langacker:2000ju}, we write the couplings of the
$Z^\prime$-boson to fermions as
\begin{eqnarray}
J_{Z^{\prime}}^{\mu}=g^{\prime}\sum_i \bar\psi_i
\gamma^{\mu}[\epsilon_i^{\psi_L}P_L+\epsilon_i^{\psi_R}P_R]\psi_i,
\label{eq:JZprime}
\end{eqnarray}
where $i$ is the family index and $\psi$ labels the fermions  and
$P_{L,R}=(1\mp\gamma_5)/2$. According to some string construction or
GUT models such as $E_6$, it is possible to have family
non-universal $Z^{\prime}$ couplings. That is, even though
$\epsilon_i^{L,R}$ are diagonal, the couplings are not family
universal. After rotating to the physical basis, FCNCs generally
appear at tree level in both left handed and right handed sectors.
Explicitly,
\begin{eqnarray}
B^{\psi_L}=V_{\psi_L}\epsilon^{\psi_L}V_{\psi_L}^{\dagger},\;\;\;\;\;
B^{\psi_R}=V_{\psi_R}\epsilon^{\psi_R}V_{\psi_R}^{\dagger}.
\end{eqnarray}
Moreover, these couplings may contain CP-violating phases beyond
that of the SM. The effective Hamiltonians describing the transition
mediated by the $Z'$ boson have the form as:
\begin{eqnarray}
 {\cal H}_{eff}^{Z'}(b\to sq\bar q)= - \frac{4 G_F}{\sqrt 2} V_{tb}V_{ts}^*
\left [ \left ( \frac{g' M_Z} {g_1 M_{Z'}} \right )^2
\frac{B_{sb}^L}{V_{tb} V_{ts}^*} (B_{qq}^L O_9 +B_{qq}^R O_7)\right
],\;\nonumber\\
{\cal H}_{eff}^{Z'}(b\to dq\bar q)= - \frac{4 G_F}{\sqrt 2}
V_{tb}V_{td}^* \left [ \left ( \frac{g' M_Z} {g_1 M_{Z'}} \right )^2
\frac{B_{db}^L}{V_{tb} V_{td}^*} (B_{qq}^L O_9 +B_{qq}^R O_7)\right
],\;
\end{eqnarray}
where $g_1=e/(\sin \theta_W \cos \theta_W)$ and $B_{ij}^{L(R)}$
denote the left (right) handed effective $Z'$ couplings of the
quarks $i$ and $j$ at the weak scale. The diagonal elements are real
due to the hermiticity of the effective Hamiltonian but the off
diagonal elements may contain effective weak phases. With the
definition
\begin{eqnarray}
 y=( \frac{g' M_Z}{g_1 M_{Z'}})^2,
\end{eqnarray} we can parameterize these coefficients as
\begin{eqnarray}
\Delta C_9^{q}(b\to s)= y \left ( \frac{B_{sb}^L
B_{qq}^L}{V_{tb}V_{ts}^*}\right )=|\xi_1^{L,q}| e^{i \phi}\;,&~~~&
\Delta C_7^{q}(b\to s)= y \left (\frac{B_{sb}^L B_{qq}^R}{V_{tb}
V_{ts}^*}\right )=|\xi_1^{R,q}|
e^{i \phi}\;,\nonumber\\
\Delta C_9^{q}(b\to d)= y \left ( \frac{B_{db}^L
B_{qq}^L}{V_{tb}V_{td}^*}\right )=|\xi_2^{L,q}| e^{i
\phi'}\;,&~~~&\Delta C_7^{q}(b\to d)= y\left (\frac{B_{db}^L
B_{qq}^R}{V_{tb} V_{td}^*}\right )=|\xi_2^{R,q}| e^{i \phi'}\;,
\end{eqnarray}
where $\phi'=\phi-\beta$, ($\phi$ is
the weak phase associated with $B_{db}^L$).

In the following discussion, we adopt $B_{uu}^{L(R)}\simeq -2
B_{dd}^{L(R)} \simeq -2B_{ss}^{L(R)} $ for convenience (note that a
possible negative sign can be accounted for by shifting $\phi$ by
$\pi$), which has been stated in detail in Ref. \cite{Barger:2009hn}.
In order to see the effect of $Z'$ boson, we have to know the values
of the $\Delta C_7$ and $\Delta C_9$ or equivalently $B_{db}^L$,
$B_{sb}^L$ and $B_{qq}^{L,R}$. Generally, one expects $g'/g_1 \sim 1
$, if both the $U(1)$ gauge groups have the same origin from some
grand unified theories, and $M_Z/M_{Z'} \sim 0.1 $ for a TeV scale
neutral $Z'$ boson, which yields $y \sim 10^{-2}$. However, in Ref.
\cite{Barger:2009hn} assuming a small mixing between $Z-Z'$ bosons
the value of $y$ is taken as $y \sim 10^{-3}$. It has been shown in
\cite{Cheung:2006tm} that the mass difference of $B_s - \bar B_s $
mixing can be explained if $|B_{sb}^L| \sim |V_{tb} V_{ts}^*| $.
Similarly, the CP asymmetry anomaly in  $B \to \phi K, \pi K $   can
be resolved if $|B_{sb}^L B_{ss}^{L,R}| \sim |V_{tb} V_{ts}^*|$. So,
we assume that
\begin{eqnarray}
|\xi_1|=|\xi_1^{L,s}|=|\xi_1^{R,s}|
=\frac{1}{2}|\xi_1^{L,u}|\in(10^{-3},10^{-2}).
\end{eqnarray}
Assuming only left handed couplings are present, the bound on FCNC
$Z'$ coupling ($B_{db}^L)$ from $B^0 - \bar B^0$ mass difference has
been obtained in \cite{Cheung:2006tm} as
\begin{eqnarray} y|{\rm
Re}(B_{db}^L)^2| < 5 \times 10^{-8},~~~~ y|{\rm Im }(B_{db}^L)^2| <
5 \times 10^{-8}\;.
\end{eqnarray}
Using $y \sim 10^{-2}$, one can obtain a more stringent bound on
$|B_{db}^L| < 10^{-3}$.
{}From these two relations one can obtain
$|B_{ss}^L| \sim 1 $. Thus, it is expected that $\xi_2^{L,R} \sim
10^{-3}$ with the CKM matrix elements considered. However, in our
analysis here we vary their values within the range $|\xi_2|\in
(10^{-3},10^{-2})$, since the major purpose of this work is searching for new
physics signal rather than obtaining acute numerical results.

\begin{figure}
\begin{center}
\includegraphics[width=9.5cm,height=6cm]{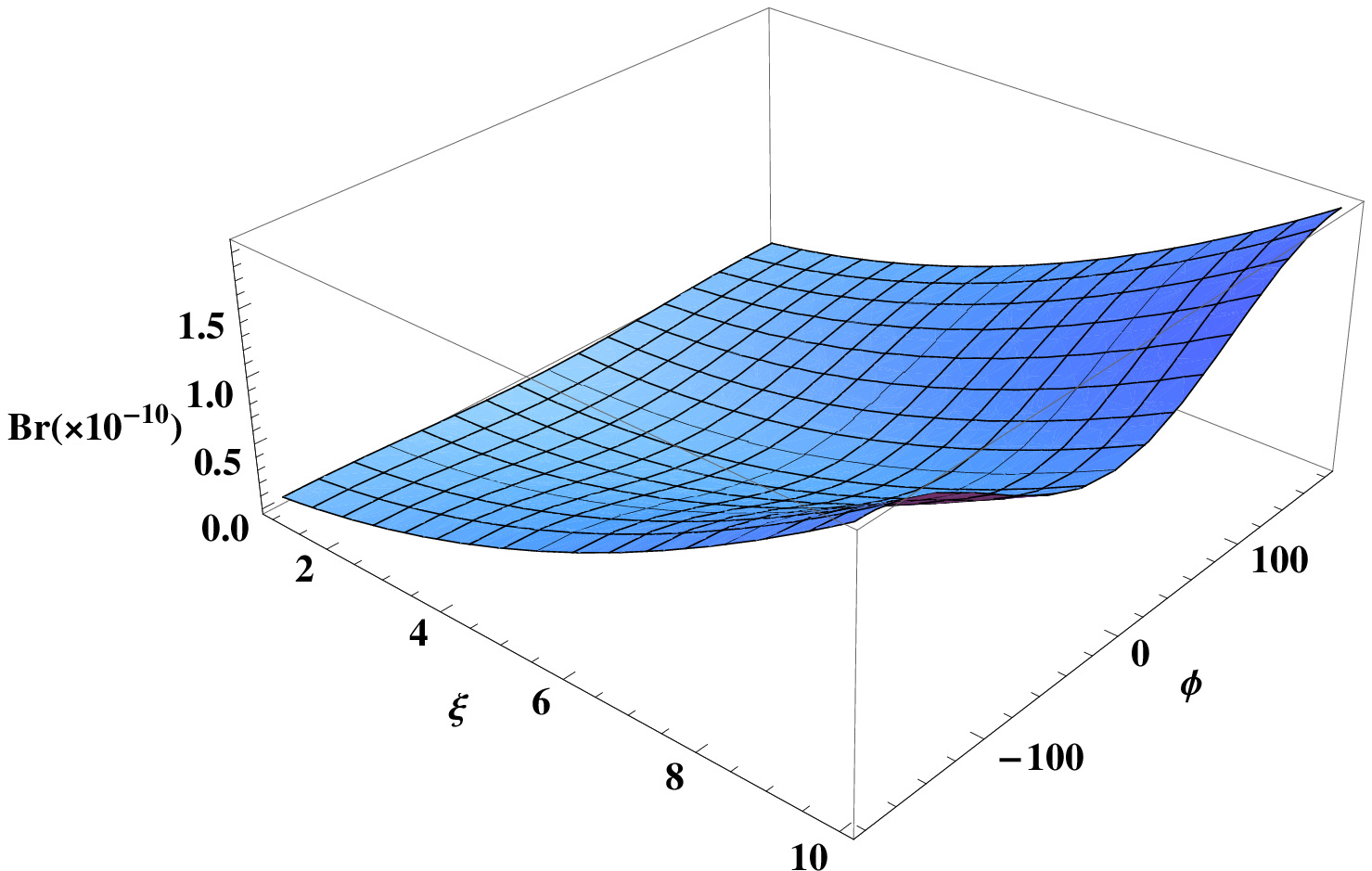}
\includegraphics[width=6cm,height=6cm]{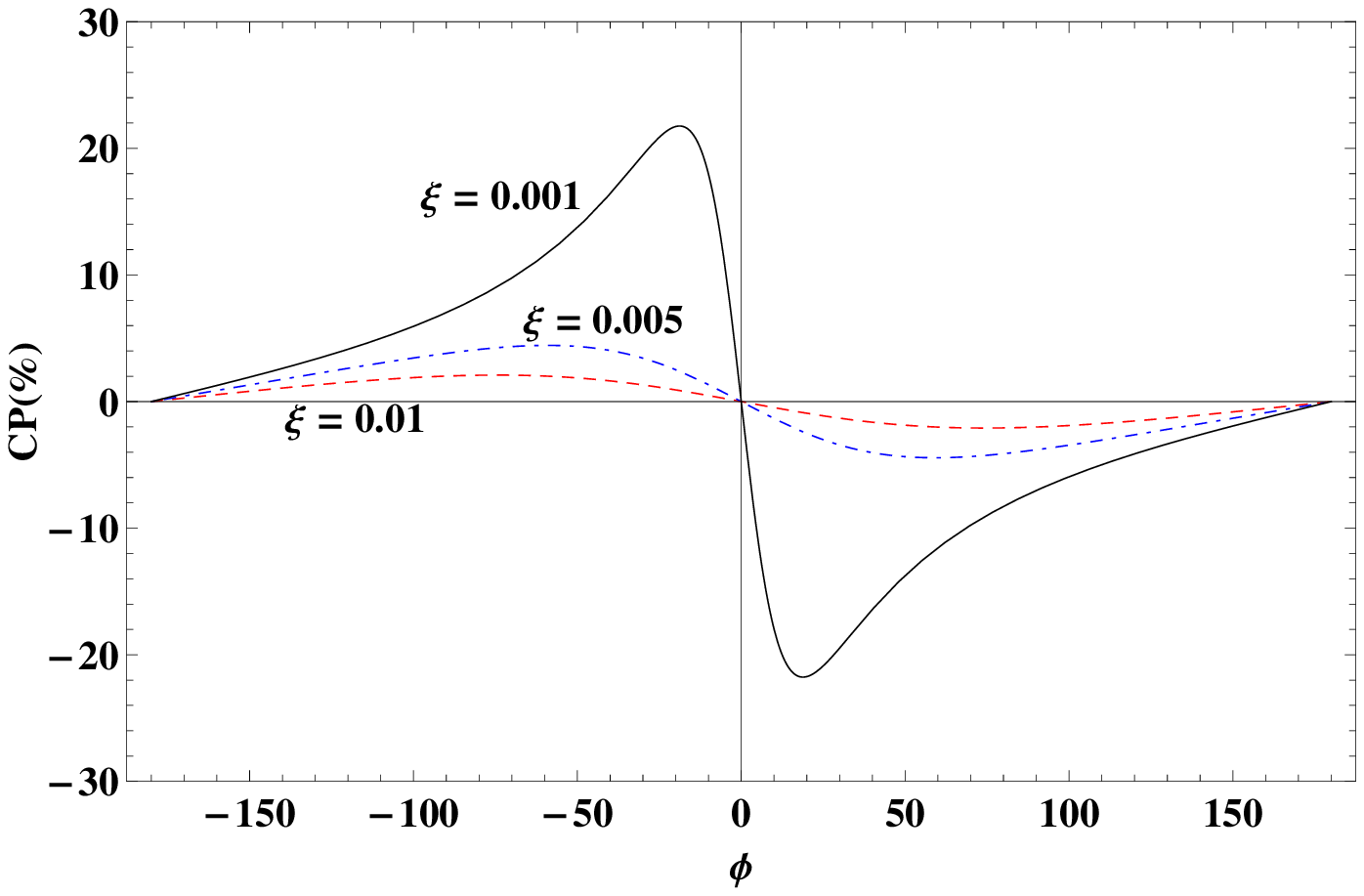}
\caption{For the decay mode $B \to \phi\gamma$, variation of the CP
averaged branching ratio (in units of $10^{-10}$) with $\xi$ (in
units of $10^{-3}$) and the new weak phase $\phi$ (left panel) and
the variation of  direct CP asymmetry
 (in \%)  with the new weak phase $\phi$ (right panel)
where the solid, dot-dashed and dashed lines correspond to
$\xi=0.001, 0.005$ and $0.01$.}\label{x1}
\end{center}
\end{figure}

\begin{figure}
\begin{center}
\includegraphics[width=8cm,height=6cm]{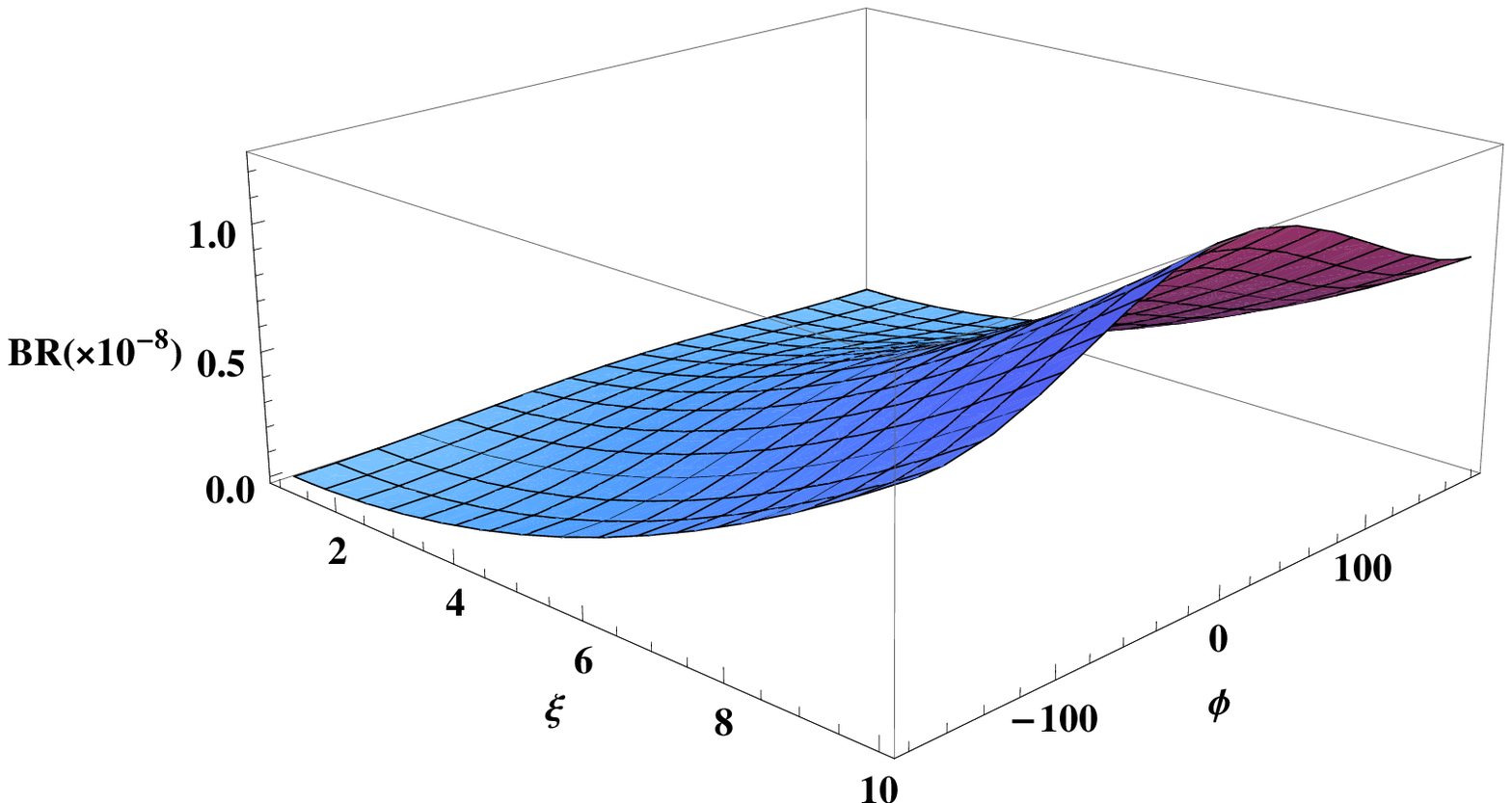}
\includegraphics[width=6cm,height=6cm]{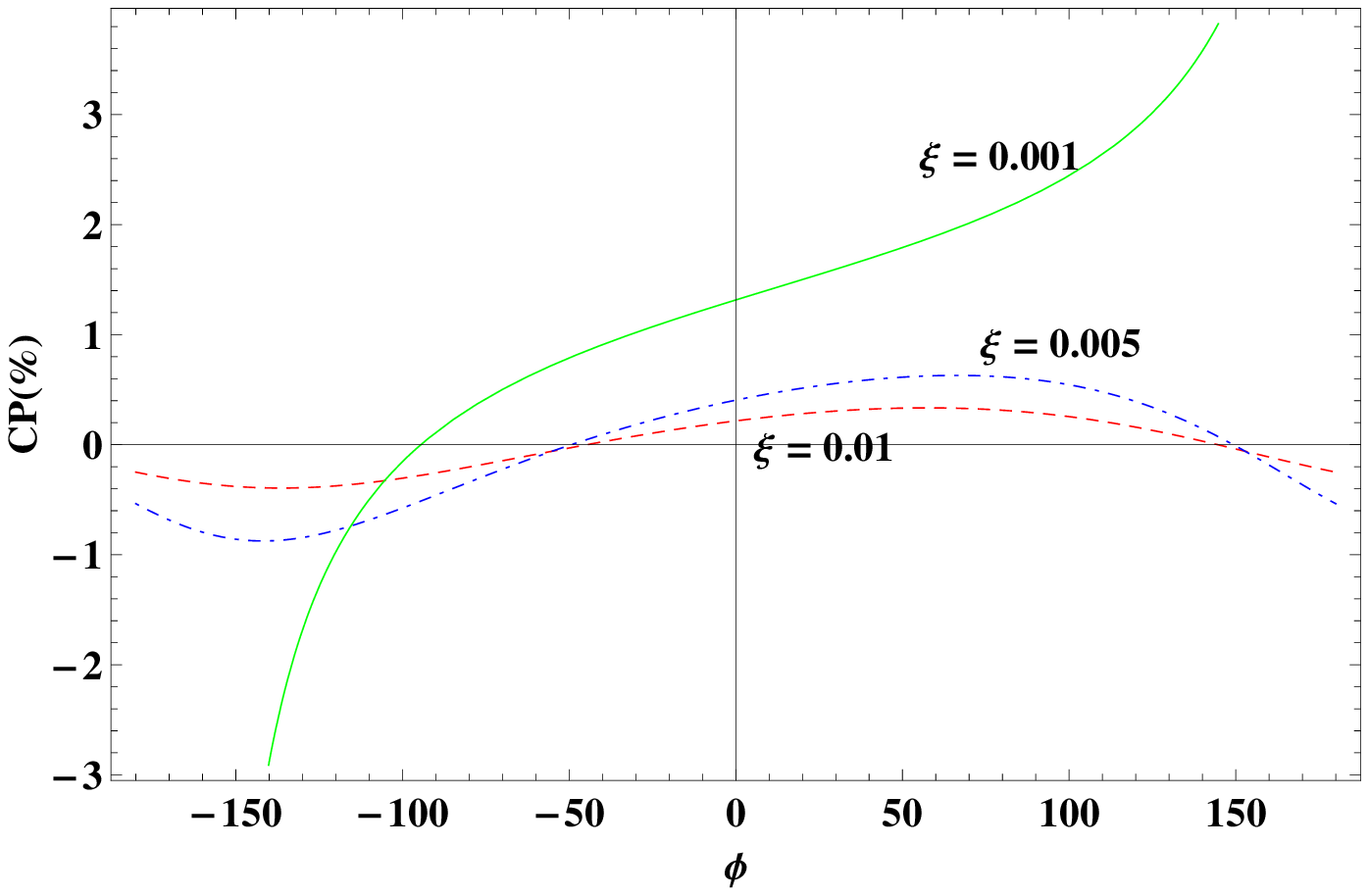}
\caption{For the decay mode $B_s \to \rho^0\gamma$, variation of the
CP averaged branching ratio (in units of $10^{-8}$) with $\xi$ (in
units of $10^{-3}$) and the new weak phase $\phi$ (left panel) and
the variation of  direct CP asymmetry
 (in \%)  with the new weak phase $\phi$ (right panel)
where the solid, dot-dashed and dashed lines correspond to
$\xi=0.001, 0.005$ and $0.01$.}\label{x2}
 \end{center}
\end{figure}

\begin{figure}
\begin{center}
\includegraphics[width=7cm,height=6cm]{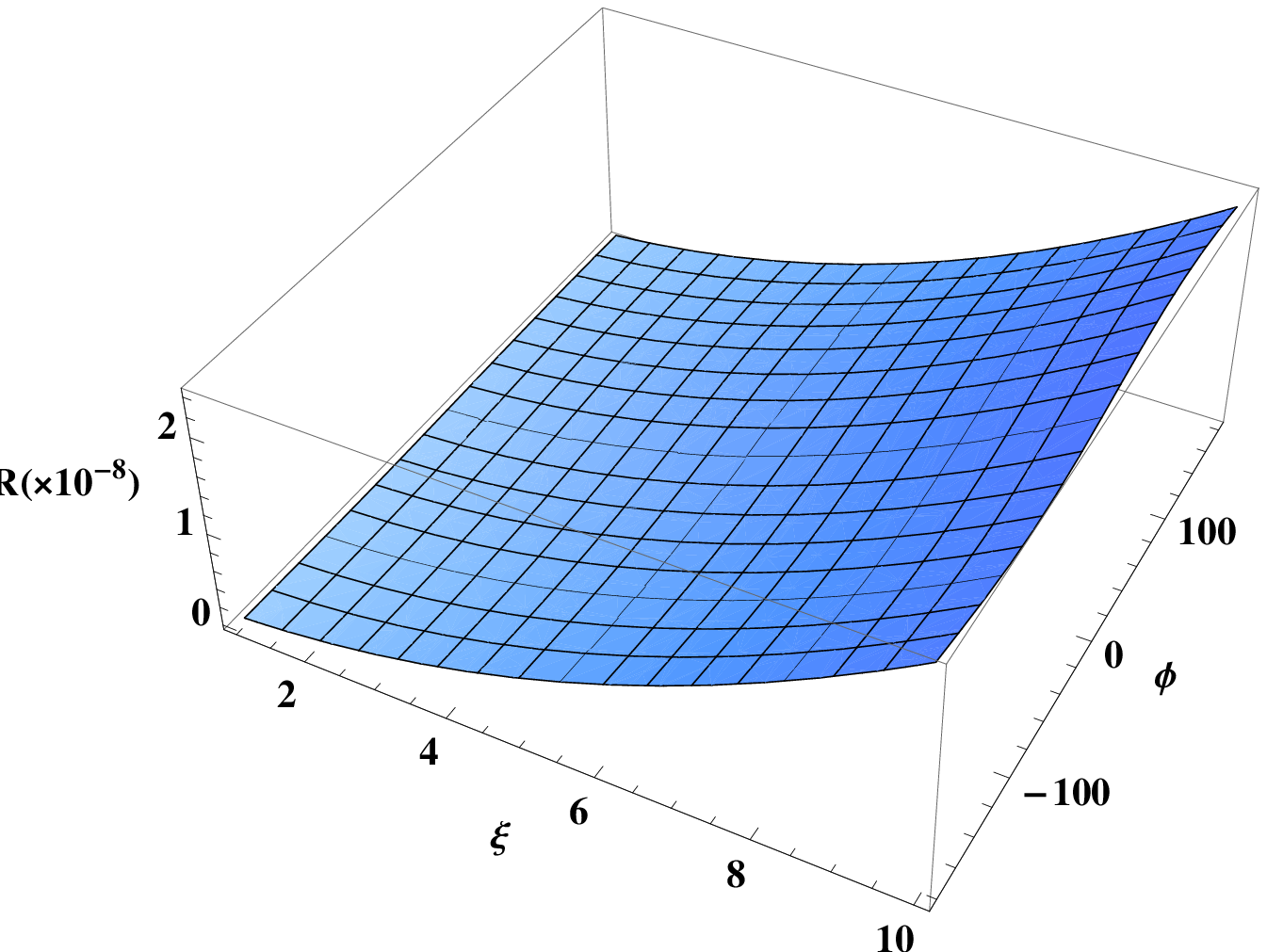}\,\,\,
\includegraphics[width=6cm,height=6cm]{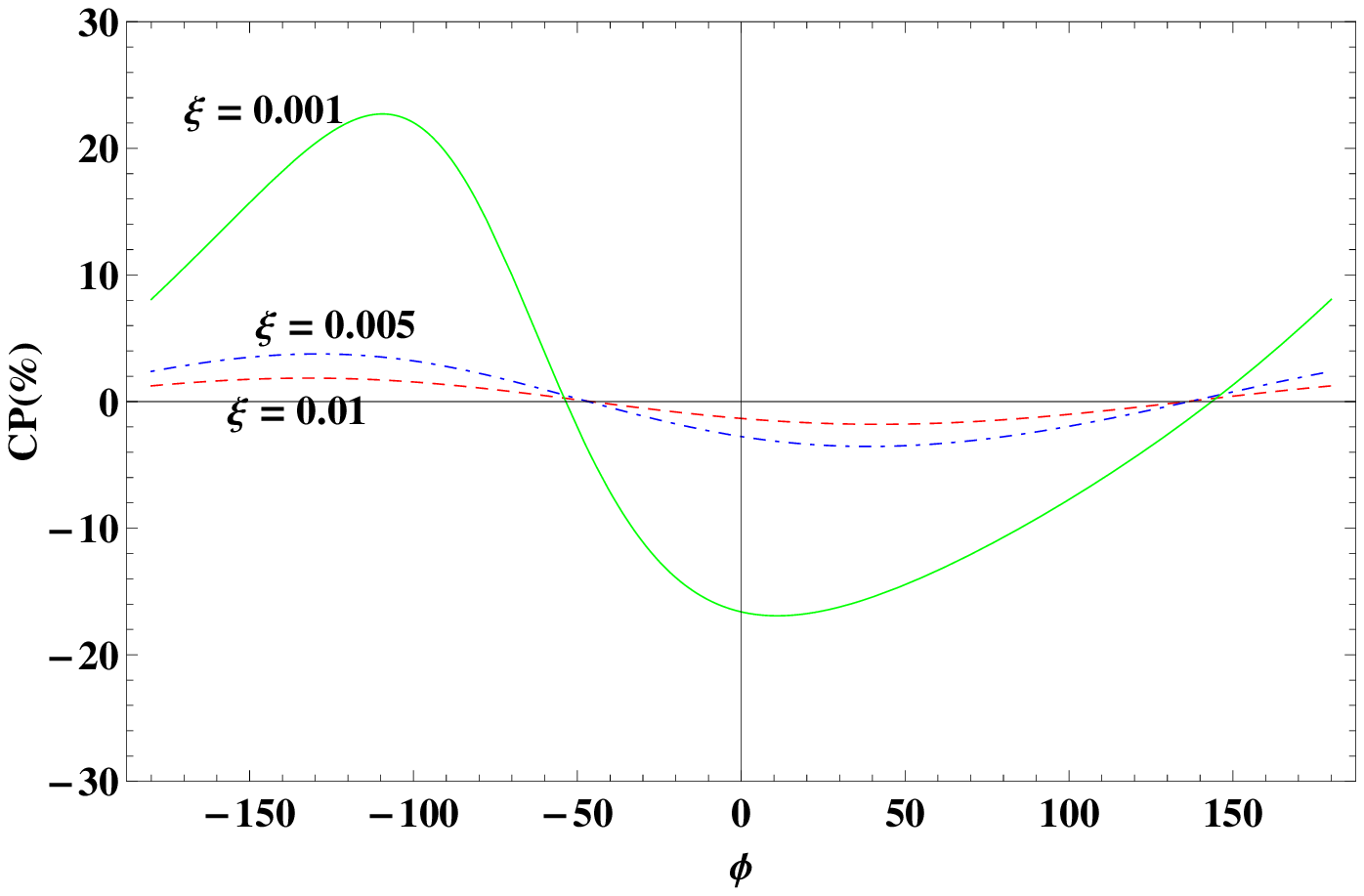}
\caption{For the decay mode $B_s \to \omega\gamma$, variation of the
CP averaged branching ratio (in units of $10^{-8}$) with $\xi$ (in
units of $10^{-3}$) and the new weak phase $\phi$ (left panel) and
the variation of  direct CP asymmetry
 (in \%)  with the new weak phase $\phi$ (right panel)
where the solid, dot-dashed and dashed lines correspond to
$\xi=0.001, 0.005$ and $0.01$.} \label{x3}
\end{center}
\end{figure}

It is noted that the other Wilson coefficients may also receive
contributions from the $Z^{\prime}$ boson through renormalization
group~(RG) evolution. With our assumption that no significant RG
running effect between $M_Z^{\prime}$ and $M_W$ scales, the RG
evolution of the modified Wilson coefficients is exactly the same as
the ones in the SM~\cite{Buras}. Using the values of these
coefficients at $m_b$ scale we can analogously obtain the new
contribution to the transition amplitude as done in the case of $Z$
boson. The  $\Delta a_i^{\prime}$ induced by the $Z^\prime $are given
as:
\begin{eqnarray}\label{a3}
\Delta a^{\prime}_2&=&\Delta C_1 +\frac{\Delta C_2}{3}
+\frac{\alpha_s}{2\pi}\frac{C_F}{N_C}\frac{f_{\rho}^{\perp}}{f_{\rho}}\Delta C_{2}F_{1},\nonumber\\
\Delta a^{\prime}_3&=&\Delta C_3 +\frac{\Delta C_4}{3}
+\frac{\alpha_s}{4\pi}\frac{C_F}{N_C}\frac{f_{\rho}^{\perp}}{f_{\rho}}\Delta C_{4}F_{1},\nonumber\\
\Delta a^{\prime}_5&=&\Delta C_5 +\frac{\Delta C_6}{3}
+\frac{\alpha_s}{4\pi}\frac{C_F}{N_C}\frac{f_{\rho}^{\perp}}{f_{\rho}}\Delta C_{6}F_{2},\nonumber\\
\Delta a^{\prime}_7&=&\Delta C_7 +\frac{\Delta C_8}{3}
+\frac{\alpha_s}{4\pi}\frac{C_F}{N_C}\frac{f_{\rho}^{\perp}}{f_{\rho}}\Delta C_{8}F_{2},\nonumber\\
\Delta a^{\prime}_9&=&\Delta C_9 +\frac{\Delta C_{10}}{3}
+\frac{\alpha_s}{4\pi}\frac{C_F}{N_C}\frac{f_{\rho}^{\perp}}{f_{\rho}}\Delta
C_{10}F_{1},
\end{eqnarray}
and the contributions of new physics can be formulated as
\begin{eqnarray}
\Delta {\cal M}_{B_s\to \rho \gamma}^{+
+}&=&i\frac{G_F}{\sqrt{2}}\sqrt{4\pi\alpha_e}F_{V}
f_{\phi}m_{\phi}M_{B}\left[V_{ub} V_{us}^*\Delta a_2^{\prime}-V_{tb}
V_{ts}^*
\left(\frac{3}{2}\Delta a_7^{\prime} +\frac{3}{2}\Delta a_9^{\prime} \right)\right],\\
\Delta{\cal M}_{B_s\to \omega \gamma}^{+
+}&=&i\frac{G_F}{2}\sqrt{4\pi\alpha_e}F_{V}
f_{\phi}m_{\phi}M_{B}\left[V_{ub} V_{us}^*\Delta a_2^{\prime}-V_{tb}
V_{ts}^* \left(2 \Delta a_3^{\prime}+2\Delta a_5^{\prime}
+\frac{1}{2}\Delta a_7^{\prime}+\frac{1}{2}\Delta a_9^{\prime} \right)\right],\\
\Delta{\cal M}_{B_d\to \phi \gamma}^{+
+}&=&i\frac{G_F}{\sqrt{2}}\sqrt{4\pi\alpha_e}F_{V}
f_{\phi}m_{\phi}M_{B}\left[-V_{tb} V_{td}^* \left(\Delta
a_3^{\prime}+\Delta a_5^{\prime}-\frac{1}{2}\Delta a_7^{\prime}
-\frac{1}{2}\Delta a_9^{\prime} \right)\right].
\end{eqnarray}
Noted that $\Delta a_i^{\prime}$s involve the new weak phase,
which may change the CP asymmetries remarkably. Now using
$|\xi_1|=|\xi_2|=\xi$ and taking the decay mode $B_s \to \rho^0
\gamma$ as an example, we also list the correction to $a_i^\prime$
from $Z^\prime$ boson in the last column of Table 2. From the table, we note that the
corresponding Wilson coefficient of electro-weak penguin is enhanced
remarkably with suitable parameter $\xi$, which may affect the
branching ratio and other observed values. In Fig. \ref{x1},
Fig. \ref{x2} and Fig. \ref{x3}, we show the variation of the CP
averaged branching ratios of $B\to \phi \gamma$, $B_s \to \rho^0
\gamma$ and $B_s\to \omega \gamma$ with $\xi$ and the new weak phase
$\phi^{(\prime)}$ (left panel) and the corresponding direct CP
violation with $\phi^{(\prime)}$ (right panel), respectively. As
anticipated, if the $\xi=0.01$, the branching ratios can be enhanced
remarkably, which can reach to
$\mathcal{O}(10^{-8})$ for $B_s \to \rho^0(\omega)\gamma$ and  $\mathcal{O}(10^{-10})$ for the
$B_d \to \phi^0\gamma$. All results
are enhanced two orders of magnitude over the predictions of the SM. From the figures, we can also argue
that there have been significant enhancements in the
branching ratios for large $\xi$, or in other words for a lighter
$Z'$ boson. In the experimental side, these results may be
inaccessible at the Belle and BaBar presently. However, it is large
enough for LHC-b and/or Super B-factories. Moreover, we find that these
decay modes may have  large CP asymmetries when $\xi=0.001$ and
suitable weak phase $\phi$. It implies that the contributions from new
physics and from the SM can be comparable, and the interference between
them leads to large asymmetries. Furthermore, future observations of
these modes could in turn help us to constrain the mass of $Z'$ boson within the model.

\section{Conclusion}
In this work, we have studied pure annihilation type radiative processes
$\overline{B}^{0}_{d}\to\phi\gamma$,
$\overline{B}^{0}_{s}\to\rho\gamma$ and
$\overline{B}^{0}_{s}\to\omega\gamma$ within the QCD factorization.
After adding the vertex corrections to the naive factorization approach, we find that the non-factorizable
contributions can enhance the branching ratio of
$\bar{B}^{0}_d\to\phi\gamma$ decays, however, the branching ratio of
$\bar{B}^{0}_{s}\to\rho\gamma$ is almost unchanged, but for
$\bar{B}^{0}_{s}\to\omega\gamma$ the branching ratio is even lowered, because the corrections to EW
penguin operators are much smaller than those to QCD penguin
operators. The smallness of these decays within the SM makes them
sensitive probes of flavor physics beyond the SM. To explore new physics
potential, we have estimated the contribution of the non-universal $Z^\prime$ model to the
decays. If $\xi=0.01$, the branching ratios can be enhanced
remarkably, and  reach to
$\mathcal{O}(10^{-8})$ for $B_s \to \rho^0(\omega)\gamma$ and  $\mathcal{O}(10^{-10})$ for the
$B_d \to \phi^0\gamma$. Moreover,
we have also predicted large CP asymmetries in suitable parameter
spaces. These results can be  tested at  the LHC-b and/or Super B-factories
in future. The observations of these modes could in turn help us
to constrain the mass of $Z'$ within the model.

\section*{Acknowledgments}

The work of C.S.K. was supported in part by Basic Science Research
Program through the NRF of Korea funded by MOEST (2009-0088395) and
in part by KOSEF through the Joint Research Program (F01-2009-
000-10031-0). The work of Y.Li was supported by the Brain Korea 21
Project and by the National Science Foundation under contract
Nos.10805037 and 10625525. Y.Li thanks Prof Y.-D Yang for valuable
discussion.

\end{document}